\documentclass[%
 reprint,
 amsmath,amssymb,
 aps,
]{revtex4-2}

\usepackage{graphicx}
\usepackage{dcolumn}
\usepackage{bm}
\usepackage[mathlines]{lineno}
\nolinenumbers\relax 
\usepackage{xcolor}
\usepackage{soul}
\usepackage[caption=false]{subfig}
\usepackage{threeparttable}
\usepackage{graphicx}
\usepackage{enumerate}
\usepackage{enumitem}
\usepackage{CJK}
\CJKnospace

\begin{document}

\preprint{APS/123-QED}

\newpage

\title{The Directionality of\\ Gravitational and Thermal Diffusive Transport \\ in Geologic Fluid Storage}

\author{Anna L. Herring}
 \affiliation{Civil and Environmental Engineering, University of Tennessee, Knoxville, TN 37996}
 \email{aherri18@utk.edu}
 
\author{Ruotong Huang \begin{CJK*}{UTF8}{gbsn}(黄若\end{CJK*}\begin{CJK*}{UTF8}{bsmi}橦) \end{CJK*}}
\author{Adrian Sheppard}

\affiliation{%
 {Materials Physics, The Australian National University, Canberra ACT 2600, Australia}
}%

\date{\today}

\begin{abstract}
Diffusive transport has implications for the long-term status of underground storage of hydrogen (H$_2$) fuel and carbon dioxide (CO$_2$), technologies which are being pursued to mitigate climate change and advance the energy transition. Once injected underground, CO$_2$ and H$_2$ will exist in multiphase fluid-water-rock systems. The partially-soluble injected fluids can flow through the porous rock in a connected plume, become disconnected and trapped as ganglia surrounded by groundwater within the storage rock pore space, and also dissolve and migrate through the aqueous phase once dissolved. Recent analyses have focused on the concentration gradients induced by differing capillary pressure between fluid ganglia which can drive diffusive transport (``Ostwald ripening"). However, studies have neglected or excessively simplified important factors; namely: the non-ideality of gases under geologic conditions, the opposing equilibrium state of dissolved CO$_2$ and H$_2$ driven by the partial molar density of dissolved solutes, and entropic and thermodiffusive effects resulting from geothermal gradients. We conduct an analysis from thermodynamic first principles and use this to provide numerical estimates for CO$_2$ and H$_2$ at conditions relevant to underground storage reservoirs. We show that while diffusive transport in isothermal systems is upwards for both gases, as indicated by previous analysis, entropic contributions to the free energy are so significant as to cause a reversal in the direction of diffusive transport in systems with geothermal gradients. For CO$_2$, even geothermal gradients less than 10$^o$C/km (far less than typical gradients of 25$^o$C/km) are sufficient to induce downwards diffusion at depths relevant to storage. Diffusive transport of H$_2$ is less affected, but still reverses direction under typical gradients, e.g. 30$^o$C/km, at a depth of 1000 m. This reversal occurs independent of the solute's thermophobicity or thermophilicity in aqueous solutions. The entropic contribution also modifies the magnitude of flux where geothermal gradients are present, with the largest diffusive fluxes estimated for CO$_2$ with a 30$^o$C/km gradient, despite the higher diffusion coefficient of H$_2$. We find a maximum flux on the order of 10$^{-13}$ mol/(cm$^2$s) for CO$_2$ in the 30$^o$C/km scenario; significantly lower than literature estimates for maximum convective fluxes in moderate to high permeability formations. Contrary to previous studies, we find that in diffusion and convection will likely work in concert -- both driving CO$_2$ downwards, and both driving H$_2$ upwards -- for conditions representative of their respective storage reservoirs.

\end{abstract}


\maketitle

\section{\label{sec:intro}Introduction}
Geologic formations underground offer high capacity, potentially long-term storage options for fluids such as waste carbon dioxide (CO$_2$) and gaseous hydrogen fuel (H$_2$), offering significant potential to mitigate climate change, provide energy storage, and accelerate the energy transition away from fossil fuels \cite{Bachu2008,Bui2018, Epelle2022, Tarkowski2022, Krevor2023SubsurfaceFuture, Yang2023}. Saline reservoirs --  porous host rock formations saturated with saline aqueous phase (``brine") -- comprise the largest resource of underground storage options \cite{USDOENETL2015}. Once injected into a saline reservoir underground, the partially soluble CO$_2$ or H$_2$ fluid (which will be referred to as `gas' in this work to distinguish if from the aqueous fluid phase) will flow through the architecture of the porous host rock along with the brine, with configurations and flow properties dictated by the local (pore-scale) capillary behavior of the rock-brine-fluid multiphase system. Because CO$_2$ or H$_2$ are typically non-wetting relative to the aqueous liquid, much of the injected gas may become ``snapped off" into small disconnected ganglia as aqueous wetting phase re-imbibes into the solid structure, and held by capillarity within pore bodies of the host rock {\cite{Roof1970,Juanes2006,Krevor2015,Singh2017}. This is known as capillary or residual trapping {\cite{IPCC2005}}. For capillary-trapped ganglia of injected fluid, it is then expected that transport will occur primarily through the aqueous phase: advection with the aqueous flow field, convection due to density gradients (e.g. \cite{Neufeld2010,Emami-Meybodi2015}), and diffusion due to concentration, gravitational, and thermal gradients (e.g. \cite{Rezk2022}).

This work focuses on the diffusive transport processes and how these may manifest in a multiphase system where concentration gradients of dissolved gas are imposed due to the presence of these residually-trapped ganglia. For partially soluble fluids in a multi-fluid porous media system, bubbles or ganglia of bulk gas may exist at different pressures \cite{Armstrong2014,Andrew2014,Herring2017FlowSandstone,Garing2017}; inducing solute concentration difference in the aqueous solvent following partitioning relationships (e.g. at ambient pressures, partitioning follows Henry's Law). For ganglia trapped within porous media, the capillary pressure difference between the ganglia and the aqueous phase ($P_{g} - P_{w}$) is reflected by the interface curvature ($\kappa$) following the Young-Laplace equation: \begin{equation}
    P_{g} - P_{w} = 2 \kappa \sigma
    \label{eq:YL}
\end{equation}
where $\sigma$ is the fluid-fluid interfacial tension. Because these ganglia will exist in relatively large vertical spans of the storage reservoir, the ganglia pressure will be related to the hydrostatic pressure gradients of the reservoir, and the partitioning relationships will be subject to both hydrostatic and geothermal gradients (temperature increasing with depth). Similarly, the final steady-state distribution of injected fluid molecules in the aqueous phase (after diffusion has acted) must also be determined as a function of the gravitational and geothermal gradients. Determination of the direction and rate of diffusive flux of injected fluid molecules from a residual state to an equilibrium state is thus non-trivial and subject to molecule-specific thermodynamic properties and behavior.

Recent work has identified ``Ostwald ripening" as a diffusive mechanism with potential to drive mass redistribution due to the varying pressure distribution (and thus concentration gradients) of injected gas ganglia. Ostwald ripening is hypothesized to drive mass transport from the high to low pressure ganglia (high to low local concentration) as the system evolves towards its ultimate equilibrium state. In the absence of hydrostatic and geothermal gradients, Ostwald ripening can drive fluid from high-pressure, high curvature bubbles to lower pressure, lower curvature bubbles. Multiple recent studies have highlighted that over long time frames, this could potentially drive injected gases to move upwards from small, isolated, capillary trapped ganglia to reconnect with the larger mobile gas plume sitting in place under the caprock \cite{DeChalendar2017Pore-scaleRocks,DeChalendar2018Pore-scaleRipening,Xu2019GravityInducedStability,Blunt2022,Li2022,Zhang2023Pore-ScaleRipening,Goodarzi2024,Yu2023,Feng2022}. For H$_2$ storage, this could be a benefit, as it would reduce the likelihood of gas loss due to capillary trapping; however, this scenario could reduce the long-term safety of CO$_2$ storage schemes, as migration of CO$_2$ into the plume will increase the capillary pressure below the caprock, causing lateral expansion of the plume and increasing the likelihood that the plume will break through the caprock. However, direct observations of diffusive transport due to Ostwald ripening are limited, and the existing analysis of potential long-term impacts is theoretical. Furthermore, the impact of geothermal gradients has been scarcely addressed and remains unresolved.
 
 Throughout the existing literature on Ostwald ripening and diffusive transport in subsurface gas storage, there are some persistent inconsistencies with respect to several important assumptions:
\begin{itemize}[topsep=4pt,itemsep=-4pt]
    \item CO$_2$ and H$_2$ do not exist as ideal gases in high pressure subsurface storage reservoirs -- both will most often be present as a non-ideal supercritical fluid. Consequently the partial molar volumes and fugacities of the gases must be considered; and the partitioning between ganglia and aqueous phase with depth is nonlinear (i.e., does not follow Henry's Law) in both cases.
    \item The effective density of dissolved H$_2$ is lower than that of pure aqueous phase; however the opposite is true for CO$_2$: the aqueous phase with CO$_2$ dissolved in it is more dense than pure aqueous phase. This indicates opposite directionality in the concentration gradient of gravity-driven thermodynamic equilibrium states for these two solutes.
    \item In subsurface environments, geothermal gradients exist along with hydrostatic gradients; this affects all the variables that affect the diffusive flux and induces transport by thermodiffusion, with recent work suggesting that geothermal effects may be much larger than the effects of buoyancy and capillarity on diffusive transport in many subsurface conditions \cite{Li2022,Coelho2023a,Coelho2023b}.  
\end{itemize}

\citet{Xu2019GravityInducedStability} and \citet{Blunt2022} provided good conceptual descriptions of the Ostwald ripening process as well as estimates of relevant timescales of fluid ganglia re-distribution due to Ostwald ripening; however, both neglected geothermal gradients and aspects of the non-ideality of the injected fluid phase. Ripening equilibration timescales have been further investigated from pore \cite{Yu2023} to Darcy \cite{Feng2022} scales, also under isothermal conditions. \citet{Li2022} provided a more complete thermodynamics-based analysis which incorporated many impacts of non-ideality and a simplified treatment of geothermal gradients for the case of CO$_2$ sequestration; but explicitly neglected the role of thermodiffusion. \citet{Coelho2023a, Coelho2023b} calculated thermodiffusion coefficients for CO$_2$ but conducted a partial analysis and incorrectly assumed that CO$_2$ thermophobicity would automatically drive it upwards under geothermal gradients.

This work seeks to extend previous work by providing a more generalized thermodynamic description of Ostwald ripening and general diffusive flux in subsurface systems, considering the above-noted factors, for the important cases of CO$_2$ and H$_2$ storage in saline reservoir formations. We incorporate the non-ideality of the gas phase from ganglia initialization to equilibrium. We argue that the Krichevsky-Kazarnovsky law \cite{Krichevsky1935} should be used to determine phase partitioning in geologic systems (rather than Henry's Law); and demonstrate how this applies to systems where gas-phase fugacity coefficients differ significantly from unity and in the presence of thermal gradients. Our presentation adds to previous analysis by making explicit the impacts of non-ideality in terms of fugacity, molar volume, partial molar volume (and thus, effective density), molar entropy, and Soret coefficients in quantifying concentration gradients and the directionality and magnitude of diffusive fluxes.

For application to the important gas storage technologies of CO$_2$ sequestration and underground hydrogen storage, we show that, in agreement with previous work, diffusion does indeed drive dissolved H$_2$ and CO$_2$ upwards to reconnect with the bulk gas-cap plume under \textit{isothermal} condition. However, we show for the first time that, for low to moderate geothermal gradients, the direction of diffusive transport is reversed -- acting downwards -- at storage-relevant depths. For CO$_2$, even small geothermal gradients, present in almost every potential storage site, will overwhelm capillary and buoyancy effects to drive CO$_2$ downwards for all CO$_2$ storage-relevant depths; in the case of H$_2$, upper regions of the reservoir favor upwards transport, but this is reversed for sufficient depths and geothermal gradients.

Our analysis is derived from thermodynamics first principles and is generalizable to porous underground storage reservoirs regardless of petrophysical details. We show that with this more complete consideration of these systems, the direction of diffusive flux is \textit{downwards} in many cases, contrary to the current body of literature \cite{DeChalendar2017Pore-scaleRocks,DeChalendar2018Pore-scaleRipening,Xu2019GravityInducedStability,Blunt2022,Li2022,Zhang2023Pore-ScaleRipening,Goodarzi2024,Feng2022,Coelho2023a, Coelho2023b}; our analysis thus completely reverses understanding of the implications for the long-term security of storage applications. In Section \ref{sec:isothermo} we present the analysis of the isothermal case (which generally supports the findings of existing literature), and then treat the geothermal gradient case in Section \ref{sec:temp}. Section \ref{sec:Kinetics of Diffusion} provides some estimates of flux under both cases.

\section{\label{sec:isothermo}Thermodynamics of the Isothermal Case}
For clarity, we refer to bulk CO$_2$ and H$_2$ as ``gas" phases (using the subscript $g$) throughout the text and in equations, despite the fact these fluids will exist as supercritical fluids at most depths of interest for geologic storage projects. The pure or bulk gas phase is distinguished from the dissolved ``solute" phase (subscript $s$), and the aqueous solvent (subscript $w$).

Chemical potential, $\mu$, is the change in Gibbs free energy of a system with respect to a change in amount of the component of interest at constant pressure and temperature, and can also be considered as partial molar Gibbs free energy. Chemical potential is a useful metric for multiphase systems because it provides a direct comparison of the component in bulk and solute form: equality of chemical potential for the component in two forms implies chemical partitioning and diffusive equilibrium. Furthermore, since chemical potential is a measure of free energy, the impacts of pressure ($P$), temperature ($T$), location in a gravitational field ($z$), and concentration in a solution ($x$) can all be incorporated directly; i.e. \(\mu = f(P,T,z,x)\). We begin by stating the general dependence of $\mu$ on pressure $P$. From the thermodynamic identity: 
\begin{equation}
\label{eq:mu_partial_pressure}
\frac{ \partial \mu}{\partial P}= V_m,
\end{equation}
where $V_m$ is the molar volume of the fluid. In contrast to liquids (including those with dissolved solutes) where $V_m$ can be considered constant over wide pressure ranges,  for gases and supercritical fluids, $V_m$ cannot be considered constant (Figure \ref{fig:Vm}). This points to the fundamental source of disequilibrium between dissolved and pure gas components as pressure increases moving deeper underground. In Section \ref{sec:temp} we consider disequilibrium caused by temperature as well as pressure gradients.

\begin{figure}
    \centering
    \includegraphics[width=1.0\columnwidth]{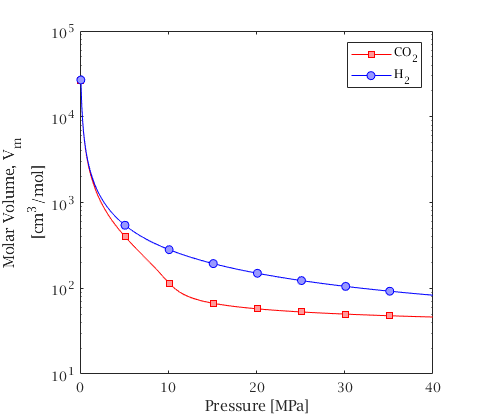}
    \caption{Molar volume values calculated via equations of state for CO$_2$ and H$_2$, under isothermal temperature of 50$^o$C. Molar volumes are identical at low pressures where both can be considered ideal gases.}
    \label{fig:Vm}
\end{figure}

 In practice, empirical Equations of State (EoS) are used to obtain values of $V_m$ as a function of pressure for the bulk gas phases. Herein, we utilize the Peng-Robinson EoS \cite{Peng1976AState} for CO$_2$; and the Abel-Noble EoS for H$_2$, using parameter values presented in \cite{Marchi2007PermeabilityPressures}; results are shown in Figure \ref{fig:Vm}. We note that although H$_2$ is a small molecule, as total system pressure increases, the molar volume of H$_2$ is significantly larger than CO$_2$, due to stronger attractive Van der Waals interactions between CO$_2$ molecules at a given pressure and temperature.

\subsection{\label{BulkAq}Bulk Gas and Aqueous Phases}

For any static fluid or fluid component in a gravitational field $g$, conservation of energy gives the dependence of chemical potential $\mu$ on height $z$ (at constant $P$, $T$) through
\begin{equation}
\label{eq:mu_depth}
\frac{\partial \mu}{\partial z} = m_m g,    
\end{equation}

where $m_m$ is the fluid component's molar mass and $g$ is the acceleration due to gravity (herein, $g$ takes a positive value in the downwards direction and height $z$ positive upwards). Note that chemical potential is sometimes defined to exclude the influence of external fields such as gravity (e.g. \cite{Li2022}), while in \cite{Kocherginsky2013}, this inclusive form of chemical potential is called the physicochemical  potential.

For a fluid column in equilibrium, the chemical potential is the same at all heights: $\mu(z) = \mu(z_o)$, leading to the hydrostatic pressure gradient \cite{Landau1969}: 

\begin{equation}
    \frac{\partial P}{\partial z} = - \rho(z) g 
    \label{eq:P_depth}
\end{equation}

For an ideal gas (``$ig$") this is integrated to give the barometric equation
\begin{equation}
     \frac{P_{ig}}{P_0} = e^{\frac{-m_m g(z-z_o)}{RT}}.
     \label{eq:P_depth_gas}
\end{equation}

\noindent Conversely, the aqueous phase can be considered an incompressible liquid with density $\rho_w$, independent of pressure, so the hydrostatic pressure $P_s$ for the aqueous solution can be calculated:
\begin{equation}
    (P_s - P_o) = - \rho_w g (z - z_o).  
    \label{eq:P_depth_aq}
\end{equation}

Here we define $z_o$ to be the height where there is mechanical equilibrium between the aqueous solution ($s$) and and bulk gas ($g$) fluid components: $P_s(z_o)=P_g(z_o)=P_o$. The difference in hydrostatic pressure gradient for the gas and aqueous solution results in mechanical disequilibrium between the fluids: $P_g(z) \neq P_s(z)$ for $z \neq z_o$. In unconstrained geometries, this leads to gravitational separation; however, in a porous medium, the two phases can coexist in mechanical equilibrium over a finite height range due to Young-Laplace pressure differences (eq. \ref{eq:YL}) arising from the non-wetting fluid forming bubble-like ganglia with positively-curved interfaces.  This results in a height-dependent capillary pressure $P_c(z) = P_g - P_s$ that can compensate for the hydrostatic pressure gradient.

The bulk solute (gas phase) will generally be less dense than the aqueous phase, and is assumed to be the non-wetting fluid in a geologic porous medium. This has been established for typical conditions and geologic materials of subsurface storage projects \cite{Krevor2015}; although shifts towards intermediate-wetting have been observed \cite{Iglauer2017, Herring2021, Herring2023}, particularly when organic carbon constituents are present \cite{Iglauer2015,Iglauer2021}. Nonetheless, we assume water-wet conditions; and therefore a ganglion of bulk gaseous (or supercritical) fluid can only exist in contact with aqueous phase above the equilibrium height, $z \geq z_o$, where $P_g > P_s$ and capillary pressure can restore mechanical equilibrium. The capillary pressure must increase as the height above $z_o$ increases, indicating smaller and smaller radii of curvature; in pore sizes typical of storage reservoirs, mechanical equilibrium can only be maintained by capillarity for a few tens of meters. This implies that as height above $z_o$ increases, the water is increasingly pushed into crevices until its presence is negligible and the pore space can be considered to be filled with bulk gas phase. At depths below the equilibrium height ($z < z_o$) the capillary pressure would need to be negative to support coexistence - this indicates that bulk gas cannot reside $z < z_o$ unless supported by a reversal of wettability.

The assumption of water-wet conditions also establishes the boundary conditions for our model: gas dissolved in the aqueous phase can migrate upwards or downwards to infinite extent. The presence of caprocks or low permeability layers may restrict movement of bulk gas, however, these water-wet layers will still permit diffusion of solute gas molecules through the aqueous phase.


We now consider dissolved gas solute within the aqueous phase; of principal interest is the behaviour of solutions with CO$_2$ or H$_2$ dissolved in water. CO$_2$ and H$_2$ are only partially soluble in water, and we treat them as dilute solutions, which greatly simplifies the analysis as the solute can be treated independently of the solvent. The dilute solution approximation is equivalent to assuming unity activity coefficients and is a standard approach for non-electrolyte solutions. For H$_2$ this assumption should be accurate given its low solubility (max 0.1 wt\%); but for CO$_2$, with a max solubility of over 10 wt\%, the activity coefficients may be substantially larger than 1, particularly for high salt concentrations \cite{Spycher2005}. Nonetheless, for lower salt concentrations (i.e 1.0 mol/kg) and the temperature and pressure ranges investigated herein, estimated CO$_2$ activity coefficients are relatively constant at around 1.25 \cite{Spycher2005}. In addition, it has been shown that the Krichevsky-Kasarnovsky equation (further explored in the next section), which is based on the dilute solution approximation, accurately predicts CO$_2$ solubility in water up to 100$^\circ$C and up to 60 MPa \cite{Carroll1992}. The dilute solution approximation is used throughout this work.

The dependence of the chemical potential of the dissolved solute phase, $\mu_s$ on pressure (other variables held constant) is the equivalent of eq. \ref{eq:mu_partial_pressure} for a solute rather than bulk phase:
\begin{equation}
\label{eq:mu_s_pressure}
\frac{\partial \mu_{s}(P)}{\partial P} = \overline{V}_m    
\end{equation}

Where $\overline{V}_m$ is the partial molar volume of the dissolved solute; i.e. the volume occupied by a mole of gas dissolved in the aqueous phase (defined formally as the increase in volume of the solution associated with addition of a mole of solute). Note that while pure gas molar volume $V_m$ is clearly not constant with pressure, the partial molar volume $\overline{V}_m$ refers to gas dissolved in the liquid phase. While CO$_2$ partial molar volume has been shown to be a function of dissolved concentration and temperature \cite{Dick1971, Parkinson1969, Garcia2001}, the data compilation of \textcite{Garcia2001} shows that $\overline{V}_{m,CO2}$ varies only between 30-40 cm$^3$/mol for temperatures from 0-100$^o$C, and the data of \cite{Parkinson1969} shows only a weak dependence on aqueous composition in the concentration range of approximately 1\% (molar percent); for simplicity, we do not model variability in $\overline{V}_{m,CO2}$ in the calculations in this work. These values imply a density higher than the aqueous solvent; this negative buoyancy is the driver for downwards convection of CO$_2$ as well as the gravitational diffusive fluxes discussed here. To our knowledge, the variation of H$_2$ partial molar volume under geologically relevant pressure, temperature and concentration is not well documented.

For dilute solutions, it is assumed that the properties of the solution are not affected by the presence of the solute, and that interactions between solute molecules can be neglected. These assumptions (which also imply that $\overline{V}_m$ is independent of concentration, and that activity coefficients are 1) lead to the standard expression for the chemical potential of a dilute solution \cite{Landau1969}: 
\begin{equation}
\label{eq:mu_s_conc}
    \mu_s(P, x) = \mu_s(P, x_o) + R T \ln \frac{x}{x_o},
\end{equation}
where  $x$ is the molar concentration, and $x_o$ is an arbitrary reference solute concentration.

To derive a relationship between $\mu$, $P$, $T$, $z$ and $x$, we begin with the definition of the total differential $d\mu(P,T,z,x)$: 
\begin{equation}
    d\mu(P, T, z, x) = \frac{\partial \mu}{\partial P} dP + \frac{\partial \mu}{\partial T} dT + \frac{\partial \mu}{\partial z} dz + \frac{\partial \mu}{\partial x} dx 
    \label{eq:diff_mu}
\end{equation}
We first apply this for dissolved gas, so that $\mu$ is the chemical potential of the solute $\mu_s$.
The partial derivatives of $\mu_s$ with respect to pressure $P$, height $z$ and concentration $x$ are obtained, respectively, from eqns. \ref{eq:mu_s_pressure},  \ref{eq:mu_depth} and from differentiating eqn. \ref{eq:mu_s_conc}; then, taking the isothermal case $dT = 0$ gives the thermodynamic identity for dilute solutions in a gravitational field:  

\begin{equation}
    d\mu_s(P, z, x) = \overline{V}_m \, dP + m_m g \,dz + \frac{RT}{x} \, dx.
    \label{eq:diff_mu_s_conc_z_v1}
\end{equation}
This can be re-written in a more convenient form, using $dP = -\rho_w g \, d z$ (from eq. \ref{eq:P_depth} for the aqueous phase), and $m_m = \overline{V}_m \rho_s $ (recall that $\rho_s$ is the effective density of the dissolved solute): 

\begin{equation}
    d\mu_s(P, z, x) = \overline{V}_m \left(\rho_s - \rho_w\right) \, g \, dz + \frac{RT}{x} dx.
    \label{eq:diff_mu_s_conc_z}
\end{equation}  

At equilibrium, $\mu_s$ does not change with height; thus, if $\rho_s \neq \rho_w$, the equilibrium concentration of dissolved solute (i.e. gas solubility) \textit{must} vary with depth, in order to compensate for the effect of a gravitational field. Thus, the equilibrium concentration profile of solute $x_e(z)$, is obtained by imposing constant chemical potential $d \mu_s(z)= 0$ in eq. \ref{eq:diff_mu_s_conc_z}:

\begin{equation}
\frac{1}{x_e} \, \frac{d x_e}{dz} = \frac{d (\ln x_e)}{dz} = - \frac{\overline{V}_m}{RT} (\rho_s - \rho_w) g 
\label{eq:diff_conc_depth}
\end{equation} 

Approximating $\overline{V}_m$ to be independent of pressure (discussed below), this equation can be integrated by defining a reference height $z_o$ such that $x(z_o) = x_o$, and integrating from $z_o$ to $z$ and $x_o$ to $x$; revealing that the gradient in concentration depends on the relative density of the dissolved phase and the water \cite{Landau1969}: 

\begin{equation}
x_e(z) = x_e(z_o)  e^{- \overline{V}_m (\rho_s - \rho_w) g ( z - z_o)/RT}
\label{eq:conc_depth_2}
\end{equation} 

Diffusive flux will act to establish this equilibrium distribution in the aqueous phase. \citet{Li2022} provide additional description and dynamic analysis of this process (often called ``sedimentation") for CO$_2$ in geologic reservoirs. 

The equilibrium distribution of solute concentration in the aqueous phase as a function of depth, calculated from equation \ref{eq:conc_depth_2} and using parameter values as indicated in Table \ref{tab:params}, is shown in solid lines in Figure \ref{fig:Conc_grads}.  As discussed above, we assume a constant partial molar volume for both CO$_2$ and H$_2$, while noting that this is an approximation only, with likely accuracy of around 25\%. There is a single unavoidable free parameter in equation \ref{eq:conc_depth_2}: the final equilibrium concentration at the reference height $x_e(z_o)$, which cannot be known \textit{a-priori} for a reservoir gas storage scheme. In this analysis, in order to estimate an upper bound on mass transport due to diffusion, we apply the assumption that $x_e(z_o)$ is equal to the solubility limit at the pressure in the aqueous phase at a depth of 2000 m. However, this assumption is highly unlikely to be achieved in the context of CO$_2$ storage; instead it is more likely that there will be \textit{no} depth at which the equilibrium concentration is equivalent to the fully saturated condition, as this would imply that enough CO$_2$ has been injected to fully saturate the formation or that downwards mass transfer due to convective dissolution has been somehow negated. 

Here we highlight an important result: because dissolved CO$_2$ is more dense than water under typical subsurface conditions, the equilibrium CO$_2$ concentration increases with depth, whereas the reverse is true for H$_2$. Some previous work has overlooked this \cite{Blunt2022} or (assuming an ideal solution) used V$_m$ in place of $\overline{V}_m$ \cite{Xu2019GravityInducedStability} leading to erroneous conclusions that the equilibrium CO$_2$ concentration gradient decreases with depth (the error in \cite{Xu2019GravityInducedStability} has been pointed out already in \cite{Li2022}).

\begin{figure}[ht]
    \includegraphics[width=1.0\columnwidth]{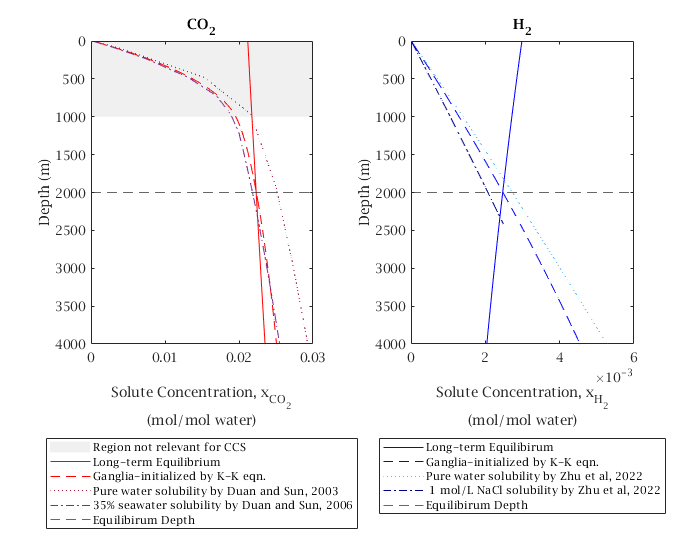} 
    \caption{Initial ganglia-initialized and equilibrium aqueous concentration profiles for CO$_2$ and H$_2$ under isothermal assumption. Ganglia-initialized concentrations are calculated from the Krichevsky-Karzarnovsky equation (eq. \ref{eq:KK}). Values interpolated from literature tabulations are included for comparison for both CO$_2$ \cite{Duan2003,Duan2006AnSO42} and H$_2$ \cite{Zhu2022} under pure water and saline conditions. Long term equilibrium shown for $\mu_i = \mu_e$ (based on the KK-concentration estimates) at $z_1=2000$ m (and depth = 2000 m). Note the difference in scale between CO$_2$ and H$_2$.}
    \label{fig:Conc_grads}
\end{figure}

Equation \ref{eq:conc_depth_2} describes the final equilibrium state of the aqueous solution in a gravitational field, while equations \ref{eq:P_depth}, \ref{eq:P_depth_gas} and \ref{eq:P_depth_aq} give the equilibrium state of a continuous gas column (the ``gas cap").  Since in both cases equilibrium derives from imposing uniform chemical potential, we can therefore also infer the equilibrium state of a multiphase fluid-porous media system  consisting of both bulk and dissolved gas; if the bulk and dissolved phases are in equilibrium at any one depth, they must be in equilibrium at all depths.  As depth increases, the pressure of the column of H$_2$ increases, and is in equilibrium with the aqueous phase which contains a \textit{decreasing} concentration of H$_2$. For CO$_2$, the increasing bulk phase pressure is in equilibrium with water containing an \textit{increasing} concentration of CO$_2$. Note that, as explained earlier, since the aqueous and gaseous phases have different densities, the phase pressures are not equal at all heights.  In the presence of thermal gradients, there is no longer diffusive equilibrium between solute and continuuous gas phase, as discussed in section \ref{sec:temp}.

\subsection{Gas Ganglia}
Now consider the non-equilibrium case where there exists a continuous aqueous liquid with dissolved solute (in a dilute solution) as the wetting phase in a porous medium, in contact with disconnected (trapped) ganglia of the solute at a height-independent capillary pressure. This scenario may arise following CO$_2$ or H$_2$ injection into an aquifer, after some re-imbibition has occurred to disconnect the gas phase. We assume that enough time has elapsed since injection for the dissolved solute to be in equilibrium with nearby trapped ganglia at the same depth, but not enough time for equilibrium between depths within the column. 

This hypothetical ``local equilibrium" scenario (also the initial condition in \textcite{Li2022}) is feasible considering that the pore-to-pore equilibration time of gas ganglia ripening is estimated to range from milliseconds to decades of years for pore sizes spanning realistic geologic porous media \cite{Yu2023}; meanwhile, equilibration across meter-long heights in a gravitational field is estimated to require several thousand years \cite{Feng2022}.  Thus, a residual, ganglia-initialized state occurs between these two equilibration times and it is our ``initial" condition, with concentration indicated as $x_i(z)$.  We note that while complete pore-to-pore equilibration in a given height may require timeframes up to years \cite{Yu2023}, our initial condition does not require complete equilibration. Instead, the largest capillary-pressure driven concentration gradients will dissipate relatively quickly, and horizontal concentration variations will be reduced to levels that are insignificant, on the meter scale, relative to gravity-induced vertical concentration differences.  The system therefore effectively attains the ganglia-initialized state well before equilibration is complete.

We first consider pressure dependence. Equilibrium between solute and ganglia implies that $\mu_s(P, x) = \mu_g(P)$ for all $P$, where $\mu_g$ is the chemical potential of the pure gas ganglia.

Therefore, integration of eq. \ref{eq:diff_mu_s_conc_z} from $P_o$ to $P$, assuming constant $\overline{V}_m$ and $T$ gives:
\begin{equation}
\label{eq:mu_g_conc_pressure}
    \mu_g(P) - \mu_g(P_o) = R T \ln \frac{x_i(P)}{x_i(P_o)} + (P - P_o) \overline{V}_m
\end{equation}

Using the fact that chemical potential $\mu_g$ and fugacity $f_g$ of the bulk phase in the ganglia are related through $\mu_g = R T \ln f_g + \text{const.}$, we see that this is the Krichevsky-Karzarnovsky equation \cite{Krichevsky1935}, a high-pressure generalisation of Henry's law:
\begin{equation}
\label{eq:KK}
    RT \ln \frac{f_g(P)}{x(P)} = R T \ln K + (P - P_{\text{vap}_w}) \overline{V}_m,
\end{equation}

where $K$ is the Henry's law coefficient, generalised for real gases:
\begin{equation*}
    K = \lim_{x \rightarrow 0} \frac{f_g}{x}
\end{equation*}
and \(P_{\text{vap}_w}\) is the vapour pressure of the solvent (water); it is insignificant for our system -- and neglected elsewhere in this work -- but necessary in general because in the limit of zero solute concentration \(x \rightarrow 0\) the ganglia are composed entirely of water vapour, so \((P - P_{\text{vap}_w}) \rightarrow 0\).  We note that the Krichevsky-Karzarnovsky equation was originally derived to describe the H$_2$-water (and N$_2$-water) system \cite{Krichevsky1935}; additionally, previous analysis of experimental data has concluded that the CO$_2$-water system is accurately modelled by the Krichevsky-Karzarnovsky equation for temperatures less than 100$^o$ C; at higher temperatures the activity of dissolved CO$_2$ must also be taken into account \cite{Carroll1992}.

With fugacity calculated from the EoSs noted above, we use eqn. \ref{eq:KK} to determine the partitioning relationship between concentration and fugacity of the pure gas phase over our considered range of depths (\ref{fig:EoS}). Fugacity coefficient is defined as the ratio of fugacity to system pressure, \(\phi = \frac{f}{P}\). Partition coefficients (ratios quantifying how a species will be distributed in two phases at equilibrium) calculated from eq. \ref{eq:KK} are presented as gas phase fugacity to aqueous solubility, with units of MPa/(mol gas/mol aq. phase), simplified to MPa.

Note that fugacity -- in this case, the equivalent pressure exhibited by the gas for phase partitioning purposes -- shows opposing behaviour for H$_2$ and CO$_2$. For H$_2$, the bulk gas phase behaves as though it is at higher pressure than the system pressure; the opposite is the case for CO$_2$. H$_2$ shows a much stronger affinity to partition into the bulk gas phase (i.e. it has a much lower dissolved concentration at the same pressure); and a more extreme variation in partitioning with depth.

\begin{figure}
    \centering
    \includegraphics[width=1.0\columnwidth]{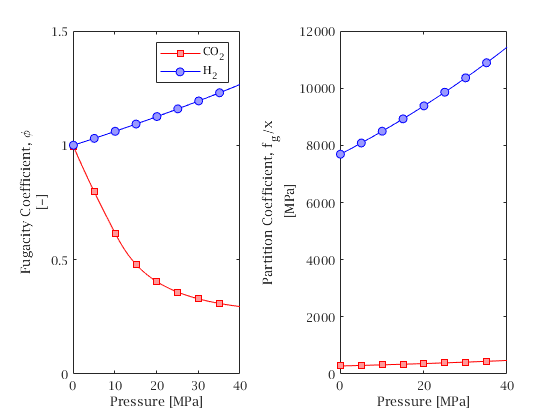}
    \caption{Calculation of gas fugacity coefficients from EoSs, and gas-solution partition coefficients from the Krichevsky-Karzarnovsky equation (eq. \ref{eq:KK}) under isothermal temperature conditions of 50$^o$C.}
    \label{fig:EoS}
\end{figure}

By applying eq. \ref{eq:diff_mu} for the chemical potential of the bulk gas phase $\mu_g$,  we can use eqs. \ref{eq:mu_partial_pressure} and \ref{eq:mu_depth} to obtain a thermodynamic identity for a column of bulk gas phase (assuming pure phase - i.e. neglecting the small fraction of water vapour)
\begin{equation}
    d\mu_g(P, z) = V_m dP + m_m g dz.
    \label{eq:diff_mu_g_z}
\end{equation}

Relating \ref{eq:diff_mu_s_conc_z} and \ref{eq:diff_mu_g_z} by imposing equilibrium between the ganglia and the dissolved solute at all depths, $\mu_s(z, x) = \mu_g(z)$, and simplifying gives the vertical concentration gradient
\begin{equation}
    RT \frac{1}{x_i}\frac{d x_i}{d z} = RT \frac{d (\ln x_i)}{dz} = - \left(V_m - \overline{V}_m \right) \rho_w g,
    \label{eq:diff_conc_depth_local}    
\end{equation}

\noindent which is determined by the density difference between the bulk and dissolved phases. In all typical scenarios, the dissolved phase will be more dense than the bulk gas phase ($V_m - \overline{V}_m > 0$), so this initial concentration gradient will increase downwards.

To illustrate the ganglia-initialized aqueous solute concentration distribution, we first estimate that ganglia pressure is 100 kPa higher than the hydrostatic pressure (to account for Laplace pressure of the gas ganglia); this is significantly higher than would generally be expected based on capillary pressure-saturation relationships for model quarry sandstones (e.g. \cite{Raeesi2014CapillaryApparatus,Herring2017FlowSandstone}) but this overestimate has no impact on our findings. Then, converting the ganglia pressure to fugacity, $f$, through the EoSs and applying the Krichevsky-Karzarnovsky derived partition coefficient, we arrive at an aqueous concentration value that would be present in the local aqueous phase equilibrated with these ganglia (Figure \ref{fig:Conc_grads}). Note that the same result is obtained by numerical integration of eq. \ref{eq:diff_conc_depth_local} using values for $V_m$ from the EoS. In Figure \ref{fig:Conc_grads} we also plot concentration values derived from interpolation of literature tabulations; including CO$_2$ solubility in pure water \cite{Duan2003}, CO$_2$ solubility in seawater-type brine \cite{Duan2006AnSO42}, and H$_2$ solubility in pure water and 1 mol/kg NaCl \cite{Zhu2022}. In the isothermal case, the concentrations estimated via eq. \ref{eq:KK} are quite similar to the literature results, despite the fact that the Krichevsky-Karzarnovsky equation includes no correction for activity.

This ``ganglia-initialized" condition represents the initial condition of this model. The residual gas saturation is not explicitly parameterized: we only assume that the aqueous phase is in equilibrium with bulk gas at hydrostatic plus capillary (100 kPa) pressures.

Comparison of the ganglia-initialized and equilibrium concentrations shows the direction of the concentration difference driving diffusive transport. Diffusion will drive the system from the concentration gradient described by eq. \ref{eq:diff_conc_depth_local} (the ganglia-initialized state) to the global equilibrium gradient of eq. \ref{eq:conc_depth_2} (the final state). Setting a reference height $z_1$ to be the point where the initial and final concentrations are the same (labeled ``Equilibirum Depth" in Figure \ref{fig:Conc_grads}); then, above $z_1$ the initial concentration is lower than the final state - there must be an influx of dissolved gas.  Below $z_1$ the initial concentration is higher than the final state - gas must be depleted to reach equilibrium. This implies upward migration of solute in all isothermal cases, even though the equilibrium concentration is increasing downwards for CO$_2$.

This conceptual description of the system in terms of concentrations is intuitive, and it is accurate in the isothermal case regardless of selection of equilibrium height $z_1$ because concentration profiles are monotonic with depth. This highlights that the final condition (i.e. the choice of the reference height $z_1$, and corresponding final concentration) is arbitrary in this case. However, we caution that this logic can only be applied when both initial and equilibrium concentration profiles are monotonic. We will show in section \ref{sec:temp} that for non-monotonic concentration gradients, the direction of diffusion cannot easily be inferred, because the concentration difference between initial and equilibrium states depends on the choice of $z_1$. Instead, it is more useful to directly interrogate the chemical potential of the system. From eq. \ref{eq:diff_mu_s_conc_z}: 

\begin{equation}
    \frac{d\mu_{si}}{dz} = (\rho_s - \rho_w) g \,\overline{V}_m + RT \frac{d}{dz} ln (x_i) 
    \label{eq:dudz1}
\end{equation}

This can be simplified by writing the concentration gradient term using eq. \ref{eq:diff_conc_depth_local} and using the fact that $\rho_s \overline{V}_m = m_m = \rho_g V_m$:

\begin{equation}
\frac{d\mu_s}{dz} = (\rho_g - \rho_w) g \, V_m 
    \label{eq:dudz}
\end{equation}

This reveals that the chemical potential gradient is controlled by the difference in density between the bulk gas and the aqueous solution, and since $\rho_g < \rho_w$ for CO$_2$ and H$_2$, the chemical potential of dissolved solute in the ganglia-initialized state decreases with $z$ (i.e. increases with depth). Under global equilibrium, \(\mu_s\) is the same everywhere and \(d\mu_s/dz = 0\) -- comparison of these two gradients demonstrates that diffusion must act upwards, as described earlier. In general, in the isothermal case: for any fluid that is less dense than the aqueous phase, the concentration in local equilibrium with trapped ganglia increases with depth more quickly than the global equilibrium concentration. Therefore Ostwald ripening will occur upwards for isothermal dilute solutions of all fluids less dense than brine.

In summary, under the assumptions of dilute solution, constant partial molar volume, and isothermal conditions; we find that:
\begin{enumerate}
    \item  In global equilibrium, $\mu_s(z) = \text{const.}$, and the dissolved concentration gradient is proportional to $-(\rho_s - \rho_w)$; thus x$_{H_2}$ and x$_{CO_2}$ have opposing gradients.
    \item For a solution in local equilibrium with trapped ganglia $\mu_{si}(z) = \mu_g(z)$; chemical potential gradient is proportional to $(\rho_s - \rho_g)$; i.e. $\mu_{si}$ increases downwards for both H$_2$ and CO$_2$, driving upwards diffusive mass transport of both H$_2$ and CO$_2$ under isothermal conditions.
\end{enumerate}

This analysis thus echos previous studies finding upwards transport due to Ostwald ripening in isothermal systems \cite{Xu2019GravityInducedStability,DeChalendar2017Pore-scaleRocks,DeChalendar2018Pore-scaleRipening,Blunt2022,Li2022,Feng2022}.

\begin{table*}
\begin{threeparttable}
\caption{Parameters used in numerical estimates}
\label{tab:params}
\centering
\begin{tabular}{l c c}
\hline\hline
\textbf{Parameter} \\
\hline\hline
Aqueous Density [kg/cm$^3$]\tnote{a}   & 1.050 $\times$ 10$^-3$ \\
Aqueous Molar Volume [cm$^3$/mol] & 18.07 \\
Acceleration of Gravity [m/s$^2$] & 9.81 \\
Isothermal Temperature [$^o$C]\tnote{a} & 50 \\
Non-isothermal Surface Temperature [$^o$C] & 25 \\
Vapor Pressure of Water [MPa] & 0.012\tnote{b} \\
\hline
\textbf{Parameter} & \textbf{CO$_2$} & \textbf{H$_2$}\\ 
\hline
Henry's Law Constant, 50$^o$C \tnote{c} [MPa] & 281 &  7683 \\
Partial Molar Volume [cm$^3$/mol] &35.1\tnote{d} & 26.7\tnote{e} \\
Molar Mass [kg/mol] & 0.044 & 0.002 \\
Diffusion Coefficient (pure water, 25$^o$C) [m$^2$/s] & $2.2\times 10^{-9}$\tnote{f} & $5.11\times 10^{-9}$\tnote{g} \\
Critical Temperature [$^o$C] & 30.978 & -239.95 \\
Critical Pressure [MPa] & 7.38 & 1.30 \\
Acentric Factor\tnote{h} [-] & 0.228 & -0.220 \\

\end{tabular}
\begin{tablenotes}
\item a. Assumed, following \cite{Blunt2022}.
\item b. \cite{Pamua2023}
\item c. Calculated from values presented in \cite{NIST} and converted to pressure units using presented values for brine.
\item d. \cite{Garcia2001}
\item e. \cite{Moore1982}
\item f. \cite{Cadogan2014}\\
\item g. \cite{TheEngineeringToolbox}
\item h. \cite{Yaws2001}
\end{tablenotes}
\end{threeparttable}
\end{table*}

\section{\label{sec:temp}Impact of Temperature Gradients}
Geologic storage will, of course, be affected by geothermal gradients. The non-isothermal case adds significant challenges to the analysis and relatively few works have analysed the impact of typical geothermal gradients. \citet{Li2022} provide an initial treatment of non-isothermal conditions; while they pointed out that this treatment requires taking into account thermodiffusion (the Soret effect), they did not include it in their models due to the scarcity of data on thermodiffusion in CO$_2$-water system; instead their model only accounted for thermal gradients by incorporating the change in solubility with depth. Their results showed thermal gradients suppress diffusive fluxes; and under certain rare conditions, have the potential to reverse their direction - i.e. for CO$_2$ to flow downwards.  As we show later, the simplified analysis in \textcite{Li2022} greatly underestimates the effect of geothermal gradients; in fact, fluxes reverse direction for low to moderate geothermal gradients (diffusive transport is downwards for both gases) and rates may increase by an order of magnitude when the entropic impacts on chemical potential are properly considered.

Following the work of \textcite{Li2022}, Coelho et al. \cite{Coelho2023a, Coelho2023b} conducted non-equilibrium molecular dynamics (NEMD) simulations  to determine the Soret coefficient of CO$_2$ in water and brine at reservoir conditions.  Their values for pure water corroborate experimental results from \textcite{Guo2018}; while there is still significant uncertainty, the effects of thermodiffusion on CO$_2$ storage can now be estimated quantitatively.  \citet{Coelho2023a, Coelho2023b} also conducted a partial analysis of CO$_2$ diffusive fluxes, incorrectly assuming that CO$_2$ thermophobicity would necessarily drive CO$_2$ upwards under geothermal gradients.  To our knowledge, Soret coefficients for H$_2$ under reservoir-relevant temperatures and pressures are still not well characterized.

There has been significant debate about whether thermodiffusion must be treated as a non-equilibrium kinetic phenomenon, or can be treated by local thermodynamic equilibrium \cite{DuhrBraun2006, Wurger2013}. \textcite{Kocherginsky2013,Kocherginsky2016,Kocherginsky2021} have made significant progress on developing local equilibrium theory of thermodiffusion and conclude that it is valid given the following assumptions \cite{Kocherginsky2021}:
\begin{enumerate}[topsep=4pt,itemsep=-4pt]
    \item transport is diffusive without hydrodynamic contributions;
    \item particle numbers $N_i$ in a volume $\delta V(z)$ under consideration are large enough ($N_i \gg \sqrt{N_i}$).
    \item local average thermodynamic variables remain meaningful; and the shortest time scale $\delta t$ is long enough so that the local equilibrium may be assumed: a local temperature $T(x)$ and concentration $c(z)$ can be defined in each volume $\delta V$.
\end{enumerate}

Condition 1 is a fundamental assumption of this work due to its focus on diffusive transport: we assume that there is no advective or convective transport within the storage reservoir. The remaining two conditions will clearly be valid for subsurface storage environments where thermal gradients are some decades of degrees per kilometer.

Therefore, we treat the non-isothermal case by assuming the reservoir consists of vertical subsections in local thermodynamic equilibrium. Return to eq. \ref{eq:diff_mu} where, rather than assuming $dT = 0$, we assume a thermal gradient $T'(z)$ such that $dT = T' dz$.  

First, considering the dissolved gas phase and using the definition of partial molar entropy $\overline{S}_m$ from the thermodynamic identity:
\[ \frac{\partial \mu_s}{\partial T} =-\overline{S}_m, \]
gives the generalization of equation \ref{eq:diff_mu_s_conc_z} for non-constant $T$: 
\begin{equation}
    d\mu_s = - \overline{V}_m (\rho_s - \rho_w) g  \,dz - \overline{S}_m T' dz + \frac{RT}{x} \, dx.
    \label{eq:diff_mu_s_conc_z_v3} 
\end{equation}

Within the above-stated assumptions we can associate the partial molar entropy with the Soret coefficient $\mathcal{\mathcal{S_T}}$ \cite{Kocherginsky2021}: 

\begin{equation}
    \mathcal{\mathcal{S_T}} = - \frac{\overline{S}_m}{RT}
\end{equation}

As mentioned above, the value of $\mathcal{S_T}$ for supercritical CO$_2$-water systems has only recently been determined, through the experimental work of \citet{Guo2018} and computational works of \citet{Coelho2023a, Coelho2023b}. These works agree on the magnitude and trend of the Soret coefficient, finding that $\mathcal{S_T}$ for CO$_2$ in pure water is positive with values between $0.01-0.03$ K$^{-1}$ at lower temperatures, and transitions to negative values in the region 370-400 K; note: a positive Soret coefficient implies a tendency to migrate to lower temperature regions. The impact of salts has been estimated by \citet{Coelho2023b}, who show a less positive $\mathcal{S_T}$, transitioning to negative values at lower $T$.
Earlier, \textcite{Windisch2012} had not found a Soret coefficient significantly different than zero, but this work had a very large uncertainty so is consistent with the values of refs \citet{Guo2018} and \citet{Coelho2023a, Coelho2023b}.

The concentration gradient in the final equilibrium state for non-constant $T$, obtained by setting $d\mu_s = 0$ (global diffusive equilibrium) in eq. \ref{eq:diff_mu_s_conc_z_v3}, is: 
\begin{equation}
\frac{d (\ln x_e)}{dz} = - \frac{\overline{V}_m}{RT} (\rho_s - \rho_w) g - \mathcal{S_T} T'.
\label{eq:diff_conc_depth_temp}
\end{equation} 
As discussed, for CO$_2$ where $\rho_s > \rho_w$, the first term is negative, leading to negative concentration gradient (increasing with depth) for the isothermal case $T'=0$.  

However, the additional term is positive at lower temperatures (where $\mathcal{S_T} > 0$), since $T$ increasing with depth implies $T'<0$.  In Figure \ref{fig:grav_thermo} we display the relative size of the gravitational and thermal contributions to eq. \ref{eq:diff_conc_depth_temp} for a range of depths, using the Soret relationship for CO$_2$ in 1 mol/kg NaCl brine from \textcite{Coelho2023b}, for isothermal conditions as well as typical geothermal gradients of 10 $^o$C/km and 30 $^o$C/km. The Soret effect does not exist in isothermal conditions, hence the 0 thermal term in the leftmost column. The gravitational term exists as a negative term under all temperature conditions; however, its magnitude remains on the order of $10^{-5}$ m$^{-1}$ -- too small to be seen - and its contribution is completely overwhelmed by the Soret effect in non-isothermal cases. We note that similar estimates are not available for H$_2$ due to the absence of published Soret coefficient data for aqueous solutions of H$_2$.

\begin{figure}
    \centering
    \includegraphics[width=1.0\columnwidth]{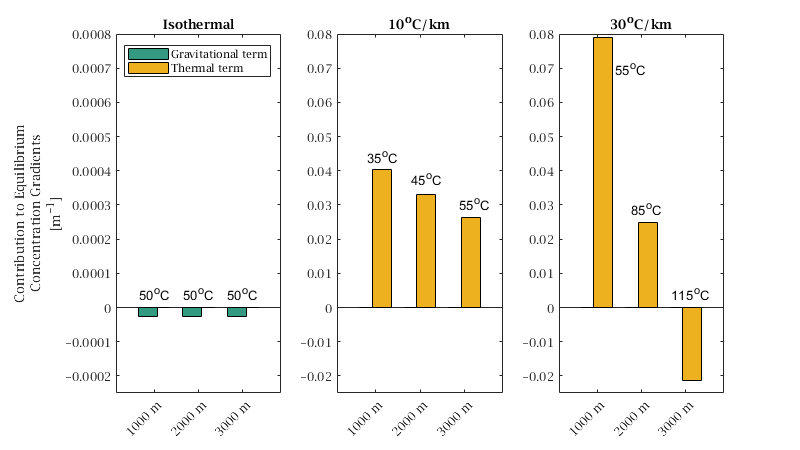}
    \caption{Contributions of gravitational and thermal terms to the gradient of natural logarithm of CO$_2$ concentration at equilibrium (eq. \ref{eq:diff_conc_depth_temp}). Note that the y-axis for the isothermal condition is an order of magnitude smaller than the cases with a geothermal gradient. The gravitational contribution is present in all cases, but too small to be visible under scenarios with a geothermal gradient.}

    \label{fig:grav_thermo}
\end{figure}

The thermodiffusion term will be sufficient to invert the concentration gradient in regions of the reservoir: equilibrium CO$_2$ concentrations will decrease with depth in middle-upper regions of reservoirs that have a typical geothermal gradient, before increasing again once the Soret coefficient transitions to negative values. While these qualitative statements can be made, quantitative estimates of the equilibrium concentration profile have very high uncertainty, since numerical integration of eq. \ref{eq:diff_conc_depth_temp} amplifies the uncertainties in $\mathcal{S_T}$. As we show below, the equilibrium concentration profile is, in fact, not significant in determining the direction of diffusive flux for either CO$_2$ or H$_2$, because it is not the gradient of \textit{concentration} that determines the direction of diffusive flux, but the gradient of \textit{chemical potential}.

For the gas phase, equation \ref{eq:diff_mu_g_z} is modified by a similar additional term for the non-isothermal case:

\begin{equation}
    d\mu_g(P, z) = V_m dP - S_m T' dz - m_m g dz.
    \label{eq:diff_mu_g_z_v3}
\end{equation}

Where $S_m$ is the molar entropy of the gas phase. 

As for the isothermal case, the ganglia-initialized (residual-state) concentration gradient in equilibrium with the gas ganglia is obtained by equating the gas and solute chemical potential from eqs. \ref{eq:diff_mu_s_conc_z_v3} and \ref{eq:diff_mu_g_z_v3}: 

\begin{equation}
    RT \frac{d (\ln x_i)}{dz} = - \left(V_m - \overline{V}_m \right) \rho_w g - (S_m - \overline{S}_m) T' 
    \label{eq:diff_conc_depth_local_T}    
\end{equation}

Differentiating eq. \ref{eq:diff_mu_s_conc_z_v3} and using eq. \ref{eq:diff_conc_depth_local_T} gives the chemical potential gradient for gas molecules in the ganglia-initialized solution:

\begin{equation}
    \frac{d \mu_{s_i}}{dz} = (\rho_g- \rho_w) g V_m - S_m T'
    \label{eq:ganglia_dudz_geotherm}
\end{equation}

This will be proportional to the diffusive flux of a ganglia-initialised system, which will be discussed further in section \ref{sec:Kinetics of Diffusion}. \\

Equation \ref{eq:ganglia_dudz_geotherm} relies only on the assumptions of dilute solution and local thermodynamic equilibrium. It is noteworthy that the diffusive flux of dissolved gas from a ganglia-initialised residual state is independent of the Soret coefficient of the dissolved gas.  Since the temperature profile is the same for the ganglia-initialised and equilibrium states, the Soret effect, which depends only on temperature, cancels out as it makes an equal contribution to the concentration profiles in both cases. With the Soret effect not playing a role, the flux is determined by how chemical potential varies with depth in the bulk gas ganglia. Sufficient contribution by the entropic term \(-S_m T'\) causes a positive gradient in $\mu_g$=$\mu_{si}$, due to the positive entropy of the bulk gas and negative $T'$. For large enough \(-S_m T'\), this leads to free energy decreasing with increasing depth because the temperature dependence of the $-TS$ term outweighs the pressure dependence in the $PV$ term. Thus, we find that both H$_2$ and CO$_2$ in the ganglia-initialized state have a tendency to migrate towards higher temperature and that this is independent of whether they are thermophobic or thermophilic.

In order to calculate the impact of thermal gradients, We obtain values for molar entropy $S_m$ for CO$_2$ and H$_2$ under our considered pressure and temperature conditions via the following steps. The molar entropy for an ideal gas ($S^{ig}$) at a given temperature ($S^{ig}_T$)is found using the Shomate equation with parameters tabulated by \cite{NIST}, calculated from data originally from \textcite{Chase1998}, and modified for pressure through \(S^{ig}=S^{ig}_T-R ln (P/P_R)\) ($P_R$ is the reduced pressure $P/P_c$ where $P_c$ is the pressure at the critical point). Deviation from the ideal gas value is found using a departure function \cite{Kyle1984,Baumann2015}, calculated using the compressibility factor and constants calculated in the Peng-Robinson EOS \cite{Peng1976AState}. Under our considered conditions, CO$_2$ molar entropy ranges from approximately 140-220 J/(mol-K), and H$_2$ from 80-110 J/(mol-K); these calculated values are consistent with tabulated data \cite{JosephHilsenrath1955CircularSteam} to approx. 2\%. Higher pressure decreases the molar entropy while higher temperature increases molar entropy; the relationship of entropy to depth is thus nontrivial and dependent on geothermal gradient (Figure \ref{fig:entropy}).

The real-gas calculated entropy values are sufficient to reverse the chemical potential gradient for CO$_2$ at relatively shallow depths, even for low geothermal gradients; for H$_2$, a gradient of 10$^o$C is insufficient to reverse the gradient, but 30$^o$C suffices (Figure \ref{fig:d_mu_dz}). Recall that positive \(d \mu_{s_i}/dz\) indicates chemical potential increasing upwards, and therefore induces downwards-driven diffusion. Larger geothermal gradients generate higher positive gradients and shift the crossover point to shallower depths. At shallow depths \textless 1000 m, the molar volume of the gas phase varies significantly, leading to large negative gradients. Upwards diffusive transport is thus favored in very shallow regions; however, these depths are generally not relevant for gas storage schemes.

\begin{figure}
    \centering
    \includegraphics[width=01.0\columnwidth]{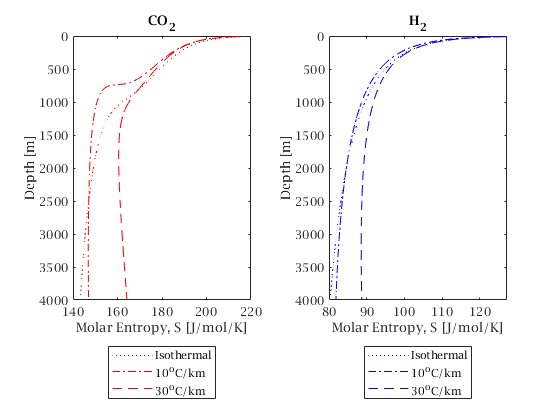}
    \caption{Variation of CO$_2$ and H$_2$ bulk molar entropy with depth for various geothermal gradients.}
    \label{fig:entropy}
\end{figure}

\begin{figure}
    \centering
    \includegraphics[width=01.0\columnwidth]{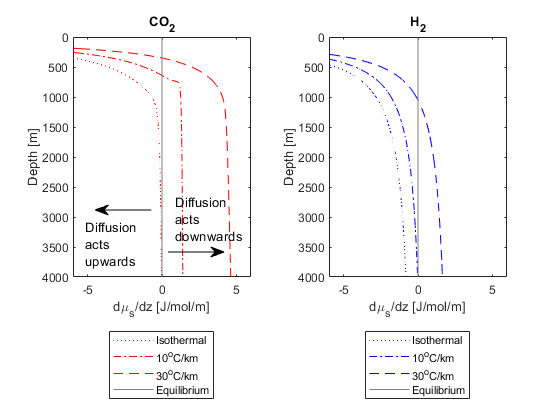}
    \caption{Variation of the vertical gradient of chemical potential  with depth for the ganglia-initialized state under different thermal assumptions, indicating the direction of diffusive flux. A positive value indicates that chemical potential is increasing with $z$ (i.e. decreasing with depth) -- resulting in downward net diffusion of dissolved gas molecules.}
    \label{fig:d_mu_dz}
\end{figure}

In Figure \ref{fig:mu_vs_depth}, we present curves of the ganglia-initialized chemical potential ($\mu_{si}$) vs. depth, calculated through numerical integration of eq. \ref{eq:ganglia_dudz_geotherm} where $\mu_s$=0 at depth = 0 (z = 4000). This corresponds to a dissolved gas concentration of essentially 0 at the surface and at (ideal mixture) solubility everywhere else; this is conceptually the same initial condition with respect to concentration as for the isothermal case, but also now includes entropic contributions to $\mu_{si}$.

With these figures, we return to the conceptual model presented in Section \ref{sec:isothermo}. At equilibirum, the chemical potential $\mu_{se}$ will be equal at all depths, i.e. present as a vertical line in Figure \ref{fig:mu_vs_depth}. The precise value at equilibrium does \textit{not} need to be determined \textit{a-prioi}, it is sufficient to know that it will be between the minimum and maximum values of $\mu_{si}$. For regions where $\mu_{si}>\mu_{se}$, the chemical potential must decrease to reach equilibrium and molecules will diffuse away; where $\mu_{si}<\mu_{se}$, there must be an influx to reach equilibrium. For curves where $\mu_{si}$ monotonically increases with depth (CO$_2$ and H$_2$ under isothermal conditions, and H$_2$ at $-T'=10^o$C), there must be diffusive transport upwards. However, for non-monotonic curves, it is possible for transport to be both upwards (for shallow depths above the inflection point) and downwards (for all depths below the inflection point). For carbon storage, reservoirs at depths below 1000 m are targeted; thus, only downwards diffusion is expected for any realistic carbon storage situation. 

\begin{figure}
    \centering
    \includegraphics[width=01.0\columnwidth]{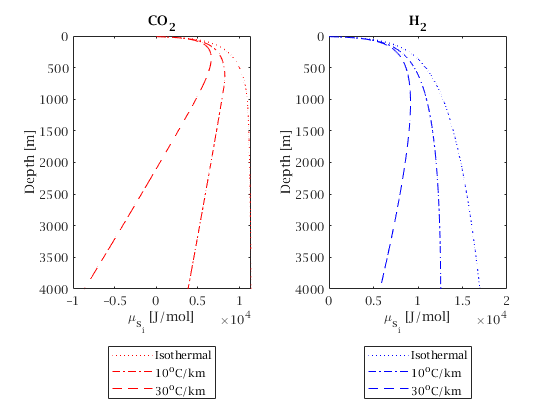}
    \caption{Variation of chemical potential $\mu_{si}$ of the ganglia-initialized system with depth. The equilibrium condition would be represented by constant $\mu$ (vertical line). The direction of diffusive flux is thus given by the sign of the gradient \(d\mu_s/dz=0\): for $\mu_s$ increasing with $z$ (decreasing with depth), diffusion acts downwards, according to eq. \ref{eq:Diff_flux}.}
    \label{fig:mu_vs_depth}
\end{figure}

It is also worth mentioning the behaviour in the upper part of the reservoir where there is continuous gas. In the presence of geothermal gradients, it is not possible for a column of gas in mechanical equilibrium ($dP = - \rho_g g dz$) to also be in diffusive equilibrium; eq. \ref{eq:diff_mu_g_z_v3} gives for the ``equilibrium" chemical potential $\mu_{g_e}$: 

\begin{equation}
    \frac{d \mu_{g_e}}{dz} = - S_m T'
    \label{eq:ganglion_dudz_geotherm}
\end{equation}

Again we see a positive gradient with $z$. This will tend to cause ex-solution at the bottom of the ganglion (lower chemical potential in the gas phase) and dissolution at the top, creating a cyclic transport of gas molecules: hydrodynamic upward flow within the gas phase and downward diffusive transport of dissolved molecules within the aqueous phase.  This is effectively the same as the process described by \textcite{Blunt2022} for the isothermal case (erroneously, as there is equilibrium in that case).  Energy for the continual motion of gas molecules comes from the continual heat flux through the reservoir.

\color{black}
\section{Kinetics of Diffusion}
\label{sec:Kinetics of Diffusion}
Flux in this diffusion-controlled system will be described by:

\begin{equation}
    J(z)=\frac{-\mathcal{D}_x(z)}{RT(z)}\frac{d\mu_{si}}{dz}
    \label{eq:Diff_flux}
\end{equation}
This expression is different from the empirically-derived Fick's first law (i.e. \(J=-\mathcal{D}\frac{dx}{dz}\)) because in this system, the gradient of chemical potential $\mu_s$ driving diffusion is not only a function of concentration $x(z)$, but also has a significant contribution due to the gravitational field and temperature gradient. 

Here we assume that the diffusion coefficient, $\mathcal{D}$ is independent of concentration and pressure, but does increase with temperature following the Stokes-Einstein equation:

\begin{equation}
    \mathcal{D}_{T(x)}=\mathcal{D}_{T=25^oC}\frac{T(x)}{298 K}\frac{\mu_{T=25^oC}}{\mu_{T(x)}}
    \label{eq:DiffusionTcorrection}
\end{equation}

Where $\mu_T$ is the dynamic viscosity of water; viscosity ratios were calculated via the correlation presented in \textcite{Kestin1978}. Diffusion coefficients were thus calculated based on the values presented in Table \ref{tab:params}; for the 10$^o$C/km and 30$^o$C/km geothermal cases, diffusion coefficients increase by a factor of approx. 2.3 and 6.3 respectively over the depth range investigated. Following the analysis of \textcite{Blunt2022}, diffusion coefficients are multiplied by an assumed porosity value of 0.2 to adjust for diffusion within porous media; this also should be an upper bound as it neglects any reduction in diffusion due to tortuosity.

The maximum diffusive flux will occur when the chemical potential gradient \(d\mu/dz\) is largest, i.e. at the ganglia-initialized state. In Figure \ref{fig:flux}, we provide estimates for flux of CO$_2$ and H$_2$ based on equations \ref{eq:Diff_flux} and \ref{eq:dudz}. Here, to provide an upper bound on flux estimates, we assign $x(z)$ to be the gas solubility in pure water at the depth-defined temperature and pressure, with $x_{CO_2}(z)$ interpolated from data reported by \textcite{Duan2003} and $x_{H_2}(z)$ interpolated from \textcite{Zhu2022}. The units of $x(z)$ in eq. \ref{eq:Diff_flux} are mol/volume, in contrast to the rest of the manuscript.

As shown in Figure \ref{fig:flux}, diffusive flux in the isothermal case is always positive (diffusion transports molecules upwards), and estimated values for CO$_2$ and H$_2$ are similar (in the region relevant to storage of both gases), on the order of $10^{-15}-10^{-14}$ mol/(cm$^2$s). As observed above, flux quickly becomes negative (diffusive transport is downwards) for CO$_2$ under even the low geothermal gradient of 10$^o$C/km; H$_2$ shows a more subtle dependence, only achieving negative fluxes below 1000 m for the 30$^o$C/km case. In the geothermal gradient case (30$^o$C), downwards CO$_2$ flux is several times larger than the flux of H$_2$; its magnitude is an order larger than the isothermal CO$_2$ case, on the order of $10^{-13}$ mol/(cm$^2$s).

This simplified analysis provides estimates of maximum flux, only reflective of the hypothetical ganglia-initialized state. Further work is needed to establish more comprehensive time-resolved kinetics, including the impacts of density-driven convection and thermal diffusive processes (e.g. \textcite{Haugen2007}).

\begin{figure}
    \centering
    \includegraphics[width=0.5\textwidth]{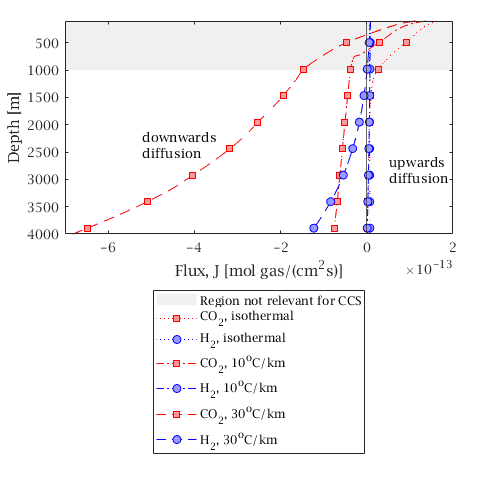}
    \caption{Diffusive flux as a function of depth for the ganglia-initialized state. Positive flux indicates upwards transport, negative flux indicates downwards.}
    \label{fig:flux}
\end{figure}

\subsection{Comparison with Mass Transport via Convective Dissolution} 
To provide an order-of-magnitude comparison, we refer to \textcite{Neufeld2010} who estimated a convective CO$_2$ flux equalling 20 kg m$^{-2}$ yr$^{-1}$ for the Sleipner site in the North sea, which is equivalent to 1.44$\times$10$^{-9}$ mol CO$_{2}$/(cm$^2$s). Compared to the upper limit of the diffusive CO$_2$ flux estimated above, the convective flux is $\approx 4$  orders of magnitude larger (depending on temperature assumption). The estimate of \textcite{Neufeld2010} represents an upper bound for storage reservoirs, due to the high permeability (5,000 mD per \textcite{Michael2010}) of the storage reservoir at Sleipner. \textcite{Emami-Meybodi2015} show that the maximum of time-resolved dissolution flux for CO$_2$ (i.e. prior to convective shutdown) decreases by $\approx$ 2 orders of magnitude when Rayleigh number (a dimensionless number describing convection, proportional to permeability) is decreased by that same factor. This indicates that even for the maximal assumptions considered above, diffusive transport of CO$_{2}$ will be a minor factor in moderate to high-permeability formations. \textcite{Li2022} provide a more thorough analysis of convective transport rates for a range of scenarios and similarly conclude that convective transport is likely to initiate and complete well before diffusive transport becomes significant, for all scenarios considered, except in low permeability ($\leq$ 1 mD) formations. For both diffusion and convection, flux is time-dependent, and this simple order-of-magnitude comparison of the maximal cases does not apply for all time periods. In particular, the low-permeability case where both convection and diffusion act on similar timescales and with similar fluxes requires further study to determine the combined effect on transport. However, it is important to note that in situations with positive geothermal gradients (i.e. all likely storage reservoirs), we predict that diffusion will drive CO$_2$ \textit{downwards}, thus increasing storage security even in the low permeability case -- in the geothermal case, diffusion and convection of CO$_2$ are not acting in competing directions.

As noted in Section \ref{sec:isothermo} above, the partial molar volume of H$_2$ dissolved in water is larger than pure water, while the opposite is true for CO$_2$; i.e. water containing dissolved CO$_2$ is more dense and water containing dissolved H$_2$ is less dense than pure water. This indicates that while convective dissolution will drive mass transport of CO$_2$ \textit{downwards}, the same process should drive H$_2$ \textit{upwards}. Estimates of upward H$_2$ flux due to this process are outside the scope of this study; however, our analysis shows that in more shallow and lower-temperature reservoirs, diffusive transport will drive hydrogen upwards, again demonstrating that convection and diffusion act in the same direction. Our analysis indicates that it would require significantly deeper and warmer reservoirs to invert the H$_2$ diffusive flux directionality. Additional study is needed to determine the comparative importance of convective and diffusive processes for application to underground hydrogen storage.

\section{Conclusions}
We have presented an analysis of diffusive transport in geologic storage scenarios. We detail the thermodynamic derivation of the phenomenon, taking into account the non-ideality of supercritical fluids (i.e. H$_2$ and CO$_2$) stored under high pressure in geologic formations. We show that under the assumption of isothermal conditions, diffusion drives injected gas molecules upwards, echoing previous analysis of the process of ``Ostwald ripening" \cite{Xu2019GravityInducedStability,DeChalendar2017Pore-scaleRocks,DeChalendar2018Pore-scaleRipening,Blunt2022,Li2022}.

However, our main contribution is in the incorporation of temperature variation via geothermal gradients. We demonstrate that with more complete consideration of entropic constributions to free energies, diffusive transport reverses direction under low to moderate geothermal gradients at storage-relevant depths; furthermore, the magnitude of flux can increase by up to two orders of magnitude. Because the entropic contribution relative to the gravitational contribution is larger for CO$_2$ than H$_2$, this reversal happens at lower temperature gradients and shallower depths for CO$_2$, with practical impacts to storage operations. Our analysis indicates that in CO$_2$ storage scenarios, diffusive transport will invariably be downwards, along with convective transport; thus both mechanisms increase the storage security of CO$_2$ storage. In H$_2$ storage systems, which are likely to be more shallow and less warm, diffusive transport will likely move gas upwards (again, in concert with convective dissolution). Under short timescales, this could make H$_2$ recovery more favorable as disconnected H$_2$ would be transported upwards to reconnect with the connected, recoverable plume under the caprock. However, this also represents a potential loss route for stored H$_2$ fuel, as diffusion upwards through the caprock is not limited by low permeability, but only the porosity and tortuosity of the confining media.  

We provide an upper estimate for flux rates, using assumptions favorable to faster diffusion. The maximum diffusive flux estimated for CO$_2$ occurs under our largest investigated geothermal gradient of 30$^o$C and is on the order of 10$^{-13}$ mol/(cm$^2$s); four orders of magnitude smaller than the downwards convective flux estimated for CO$_2$ under high-permeability conditions by \textcite{Neufeld2010}. We note that under low permeability conditions and for later times, the fluxes of convection and diffusion may become comparable. Regardless, our analysis indicates that for all typical CO$_2$ storage scenarios, regardless of permeability, both diffusion and convection act in concert to drive CO$_2$ downwards.

Our analysis has relaxed many of the assumptions prior studies have made; particularly with respect to the isothermal temperature assumption and the ideality of the gas (supercritical) phases. However, the analysis does employ two key assumptions: (1) dilute solutions, implying unity activity coefficients, and (2) local thermodynamic equilibrium in non-isothermal cases. For CO$_2$ (with a maximum solubility of over 10 wt\%) activity coefficients may be substantially larger than 1, particularly for high salt concentrations \cite{Spycher2005}; the impact of these effects on our results are worthy of further investigation. We have also used a single representative value for partial molar volume in all our calculations. At depths greater than those considered here, it is possible for CO$_2$ partial molar volume to increase to a point that the effective density of the solution is no longer greater than pure aqueous phase density, at which point the gravitational driver may reverse. Our results show that this is unlikely to impact diffusive transport of since the entropic contributions overwhelm the gravitational contributions in all storage-relevant conditions for  CO$_2$. However, this density inversion at extreme depths will have a major overall impact as it will cause a reversal of convective fluxes, likely creating a barrier to convection.

Under these assumptions, we find that (contrary to previous studies) diffusion and convection will tend to work in concert - both driving CO$_2$ downwards, and both driving H$_2$ upwards - for conditions representative of their respective storage reservoirs (i.e CO$_2$ in deeper reservoirs, and H$_2$ in more shallow formations). While still slow, diffusive transport is thus predicted to be beneficial for carbon storage; for hydrogen storage, upwards diffusion may enable re-connection of capillary trapped ganglia, but may also represent a mechanism for gas leakage.

%

\end{document}